\documentstyle{article}
\def\be{\begin{equation}}
\def\ba{\begin{eqnarray}}
\def\ea{\end{eqnarray}}   

\def\ee{\end{equation}}
\def\to{\rightarrow}

\begin{document}
\onecolumn
\begin{titlepage}
\begin{center}
{\Large \bf Pair Production of Topological anti de Sitter Black Holes } \\  \vspace{2cm}
R.B. Mann\footnotemark\footnotetext{email: 
mann@avatar.uwaterloo.ca}\\
\vspace{1cm}
Dept. of Physics,
University of Waterloo
Waterloo, ONT N2L 3G1, Canada\\
\vspace{2cm}
\vspace{2cm}
\today\\
\end{center}
\begin{abstract}
The pair creation of black holes with event horizons of non-trivial topology is described.
The spacetimes are all limiting cases of the cosmological $C$ metric. They are
generalizations of the $(2+1)$ dimensional black hole and have 
asymptotically anti de Sitter behaviour.  Domain walls instantons can mediate their
pair creation for a wide range of mass and charge.
\end{abstract}
\end{titlepage}

Pair creation of black holes continues to afford us interesting insights 
into quantum gravity and the relationship between entropy
and the number of quantum states of a black hole. It is
a tunnelling process in which the mass-energy of the 
created pair of black holes is balanced by their negative potential energy
in some  background field, such as that of an electromagnetic field 
\cite{dgkt}, a positive cosmological constant \cite{cosern},
a cosmic string \cite{strpair} or a domain wall \cite{dompair}. The
amplitude for the process is approximated by $e^{-I_i}$, where
$I_i$ is the action of the relevant instanton {\it i.e.} a Euclidean
solution to the field equations which interpolates between the states
before and after a pair of black holes is produced.  

The $C$ metric solution of the Einstein-Maxwell equations
may be interpreted as describing two oppositely-charged black holes
undergoing uniform acceleration \cite{Cmetric}. 
It contains conical singularities, which
in general cannot be eliminated at both poles. These singularities are
interpreted as representing ``rods'' or ``strings'' which provide the
force necessary to accelerate the black holes. Removal of these
singularities is generally obtained by adding additional forms of 
stress-energy which generate the background field required
to provide the necessary accelerating force. 

The purpose of this paper is to point out that the cosmological
$C$-metrics \cite{gen} contain a rich array of Euclidean instantons that
mediate pair production of black holes whose topology is of arbitrary
genus.  The genus zero solution corresponds to the set of 
Reissner-Nordstr\"om de Sitter instantons studied previously in
the context of cosmological black hole pair production \cite{cosern}.
The higher genus solutions are asymptotically anti-de Sitter,
and correspond to instantons that are 
4 dimensional generalizations of the 3 dimensional black hole \cite{btz}.
Pair production of these black holes can take place in the presence of
domain walls whose topology is the same as that of the produced black
hole pairs.

The cosmological charged $C$-metric solution is \cite{gen}
\begin{equation} \label{ccmet}
ds^2 = \frac{1}{A^2 (x-y)^2} \left[ H(y) dt^2 - H^{-1}(y) dy^2 +
G^{-1}(x) dx^2 + G(x) d\varphi^2 \right],
\end{equation}
where 
\begin{equation}\label{GHdef}
H(z) = a  - bz^2 -2mAz^3 -q^2A^2z^4 = G(z)- \frac{k l^2}{A^2}
\end{equation}
$\Lambda=3k/l^2$ being the
cosmological constant and $A$ the acceleration parameter
and $k=\pm 1$.
The gauge field is
\begin{equation} \label{gauges}
F_M = -q dx \wedge d\varphi, \qquad   F_E = -q dt \wedge dy,
\end{equation}
in the magnetic and electric cases respectively.  Under the
coordinate transformation $y=x-1/Ar$, $t=Au-\int^y dz/H(z)$, the metric 
(\ref{ccmet}) may also be written as
\begin{equation} \label{ccmetu}
ds^2 =  H(x-\frac{1}{Ar})A^2r^2 du^2 - 2dudr -2Ar^2dudx  +
r^2\left(G^{-1}(x) dx^2 + G(x) d\varphi^2 \right) \quad  .
\end{equation}
The electric field becomes $F_E = -q du \wedge (Adx +\frac{dr}{r^2})$
and the magnetic field is unchanged. 

The inner, outer and acceleration horizons are given by the first three
roots of $H$ in ascending order respectively. The coordinate 
$r\in (0,\infty)$, whereas $x$ lies between either the
smallest $(x_1,x_2)$ or largest $(x_3,x_4)$ pair of roots of $G$ so 
that the metric (\ref{ccmetu}) has proper signature.  

Removal of conical singularities in the $(x,\varphi)$ sector implies
that 
\begin{equation} \label{nostrut}
G'(x_3) =- G'(x_4),
\end{equation}
with $\varphi$ periodically identified with period $\Delta \varphi =
4\pi/ |G'(x_4)|$. This condition can only be satisfied if $x_3=x_4$.
A similar analysis for the smallest pair of roots implies that
conical singularities can be removed if $x_1=x_2$. Conical singularities
may also be avoided if $x_2=x_3$, in which case the point $x=x_3$ is an 
infinite proper distance from any allowed value 
of $x$.

Hence removal of conical singularities in the cosmological $C$ metric
implies that $G(x)$ must have a double root. This apparently
implies that the $(x,\varphi)$ sector shrinks to a point, but this
is just a poor choice of coordinates. The proper distance between
any adjacent pair $(x_i,x_{i+1})$ of roots of $G$ is actually finite,
as can be seen by setting $\varphi = \phi/\epsilon$ and
and $x = \epsilon f(\lambda)$, where the roots 
$x_i = \epsilon f(\lambda_i)$ and
$x_{i+1} = \epsilon f(\lambda_{i+1})$ coincide 
as $\epsilon \to 0$.  The parameters of $G$ must then be chosen so
that it has a double root at $x=x_i=x_{i+1}$ as $\epsilon\to 0$. This
constrains the acceleration parameter $A$ in terms of $m$, $q$ and $l$.

For all possible pairs of degenerate roots the metric (\ref{ccmetu})
becomes, as $\epsilon \to 0$,
\begin{equation}\label{rnbmet}
ds^2 = -V(r) dT^2 + \frac{dr^2}{V(r)} 
+ r^2\left(d\lambda^2 + s^2(\lambda)d\phi^2\right) ,
\end{equation}
where a coordinate transformation on $u$ has been performed, 
\begin{equation}\label{vmet}
V(r) = -\frac{k}{l^2}r^2 + b -\frac{2m}{r} + \frac{q^2}{r^2}  
\end{equation}
and $\phi$ has period $2\pi$.

If the largest two roots of $H$ are degenerate then $b>0$,
and the class of metrics obtained are of the 
Reissner-Nordstr\"om (anti)-de Sitter (RN(a)dS)
type, of mass $m$ and charge $q$. However there is a surprise 
in that the parameter $b$ is completely arbitrary. 
A simple rescaling of 
parameters and coordinates allows $b$ to be set to unit magnitude 
without loss of generality if it is nonzero. If the middle two roots 
of $H$ are degenerate, then $b=-1$, and if the largest three roots are
degenerate then $b=0$. There are therefore three possible forms for
the function $s(\lambda)$:
\begin{eqnarray}
&b=1, k=\pm 1& \quad s(\lambda) = \sin(\lambda)
\label{bm1k1} \\
&b=0, k=-1& \quad s(\lambda) = 1
\label{b0k1} \\
&b=-1, k=-1& \quad s(\lambda) = \sinh(\lambda)
\label{b1km1} 
\end{eqnarray}

It is easily checked that all of these spacetimes satisfy
the Einstein-Maxwell equations with cosmological constant.
The gauge field becomes 
\begin{equation}\label{emagform}
F_M = q s(\lambda) d\lambda \wedge d\phi \quad
F_E = -\frac{q}{r^2} dT \wedge dr
\end{equation}
in the magnetic and electric cases respectively.

The regularity requirements for the $C$ metric in
the $b=1$ case for positive $\Lambda$ have been discussed 
previously \cite{cosern}.
The other spacetimes, however, have been overlooked in
previous studies.  The $b=1$, $\Lambda<0$ case is simply
Reissner-Nordstr\"om anti de Sitter spacetime. 

The two remaining
spacetimes all have $\Lambda <0$ and $b<0$. Their
$q=m=0$ versions were studied recently by Aminneborg {\it et.al.}
\cite{amin} who showed that they  can be understood as 
four-dimensional analogues of the three-dimensional black hole
\cite{btz}, by compactifying the $(\lambda,\phi)$ sector. For the
metrics derived here this construction can also be carried out.
For $b=1$ the topology of this sector is the 2-sphere.
For $b=0$, the coordinate $\lambda$ may be identified, and the
$(\lambda,\phi)$ sector is a torus, whose unit area shall be chosen to
be $4\pi$ by identifying the $lambda$-coordinate with period 2.
For $b=-1$, the identifications may be carried out by mapping
the $(\lambda,\phi)$ sector to the Poincar\'e disk under the transformation
$\rho = \tanh(\lambda/2)$, yielding
\begin{equation}\label{pdisk}
d\lambda^2 + s^2(\lambda)d\phi^2
= \frac{1}{(1-\rho^2)^2}\left(d\rho^2 + \rho^2 d\phi^2\right)
\end{equation}
where $ 0\leq \rho < 1$.  The Poincar\'e disk has an
isomorphism group SO(2,1). Identifying points on
the disk under any discrete subgroup of SO(2,1) yields a compact 
two-dimensional space of negative curvature, which necessarily has
genus $g \geq 2$. The unit area of such surfaces is $4\pi(g-1)$.
These spaces may be constructed
by symmetrically placing a polygon of $4g$ sides
at the center of the Poincare
disk and identifying opposite sides. The edges of the polygon are
geodesics of the Poincar\'e disk; these are circles
whose extensions are orthogonal to the disk boundary. The simplest
case is the octagon with $g=2$.

The $b\leq 0$
constructions hold for all values of $r$ and $T$ in (\ref{rnbmet}).
An analysis of the behaviour of $V(r)$ in (\ref{vmet}) indicates that
in this case it has at most two roots, corresponding to an inner
and outer horizon, as with the usual RNadS metric. For $b=0$,
there will be two horizons, provided 
\begin{equation}\label{b0hor}
27\,l^{2}\,{\it m}^{4} \geq 16\,Q^{6}
\end{equation}
with the extremal case saturating the inequality. For nonzero $b$
there will be event horizons provided
\begin{equation}\label{bhor}
m^2 \leq 
  {\displaystyle \frac {l^2}{27}} \,{\displaystyle \frac { 16 -
24\,e^{2}b -16b\sqrt{1 - e^{2}b}\,e^{2} + 6b^2\,e^{4} 
+ 16\,\sqrt{1 - e^{2}b}}{e^{6}}} 
\end{equation}
where $e = \frac{2\sqrt{2}q}{3m}$. If $b=0$ (the genus 1 case)
then the range of e is from 0 to 1.  Analysis of (\ref{bhor}) in
this case indicates that event horizons can (but need not)
exist provided $q<m$. If $b=-1$ then there is no
(obvious) upper limit on e, and event horizons can exist for arbitrarily
large values of $q$ relative to $m$.

The topology of
the outer event horizon is $H^2_g$, where $H^2_g$ is a two-dimensional
surface of genus $g$. The entire spacetime has topology $R^2\times
H^2_g$. An analysis of the quasilocal mass \cite{bjm} and charge contained 
within a surface of topology $H_g$ at a fixed value of
the coordinate $r$ centered about the origin indicates,
in the limit of large $r$,
that $q(|g-1|+\delta_{g,1})$ and $m(|g-1|+\delta_{g,1})$
are the conserved charge and mass parameters
of the black hole.

Pair production of these black holes may be achieved using the
domain wall construction of ref. \cite{dompair}.  The topology
of the Riemannian section is   $R^2\times H^2_g$, where the
$R^2$ factor is like a bell.  Two copies of this
manifold may be matched together at a radius $r$ determined by the
matching condition \cite{His,Aur}
\begin{equation}\label{match}
\sqrt{V(r) - \dot{r}^2} = 2\pi\sigma
\end{equation}
where $\sigma$ is the energy per unit area of the domain wall, whose
topology is $S^1\times H^2_g$, and the overdot refers to the derivative
with respect to Euclidean proper time.
The Riemannian section is two bells glued together along their
open ends at a ridge; it
has topology $S^2\times H^2_g$  
and corresponds to static a domain wall configuration with
two surfaces at which the Killing field $\frac{d}{d\tau}$
vanishes, where $\tau$ is the Euclidean time parameter.

Equation (\ref{match}) may be interpreted
as the equation describing the motion of a fictitious particle
in a potential $v = V - (2\pi\sigma r)^2$. 
Static solutions,  which have energy zero, may
be obtained by solving (\ref{match}) under the condition 
$\partial{v}/\partial{r} = 0$.
Non-static solutions are obtained by matching the period 
of the RNadS black hole with an integer multiple of 
\begin{equation}\label{period}
\beta_w = \oint_{rmin}^{rmax} d\tau 
=  \oint_{rmin}^{rmax} \frac{dr}{\sqrt{V(V-(2\pi\sigma r)^2)}}
\end{equation}
which is the period of the domain wall evolution between the
extrema ${rmin}$ and $rmax$ where $\dot{r}$ vanishes.

For simplicity, I shall consider static solutions;
details including non-static solutions 
shall appear in a forthcoming paper \cite{forth}.
These occur when
\begin{equation}\label{matchr}
r_s =  \frac{3mb  + \sqrt{1- 8e^2b} }{2}
\end{equation}
if $b\neq 0$ and at $r_s= \frac{2 q^2}{3m}$ if  $b=0$.
Note that for $b<0$ there is no solution with zero charge.
The squared mass of the created black holes is 
\begin{equation}
m^2 = \frac {1}{54}\left[ 36 q^2 + 
\frac{b}{4\,\pi ^{2}\,\sigma ^{2}\,l^{2}-1}
\left(l^3 + \sqrt{(l^2+12 q^2(1-4\,\pi ^{2}\,\sigma ^{2}\,l^{2})^3}
\right)\right]
\end{equation}
for nonzero $b$ and  by 
\begin{equation}
m^2= \frac {4 q^3}{3\sqrt{3}l}\sqrt{4\,\pi ^{2}\,\sigma ^{2}\,l^{2}-1}
\end{equation}
if $b=0$. 

The Euclidean action for these instantons is
\begin{equation}\label{act1}
I = \int d^4x \sqrt{g} \left(-\frac{R}{16\pi}  
+ \frac{F^2}{16\pi} + {\cal L}_c +{\cal L}_d
\right)
\end{equation}
where ${\cal L}_c$ is the cosmological Lagrangian and ${\cal L}_d$ the
domain-wall Lagrangian. The former may be taken to be that
of the squared field strength of a 3-form or simply the
constant $3\frac{k}{8\pi l^2}$. The domain wall Lagrangian can be
that of a membrane current coupling to the 3-form
\cite{Aur} or that of a scalar field $\Phi$ whose potential 
${\cal V}(\Phi)$ is
everywhere positive \cite{dompair} (and so its Euclidean action is
always negative). There are no boundary terms because the instantons
considered here are compact and without boundary.

Using the Einstein field equations (\ref{act1}) becomes
\begin{equation}\label{act2}
I = \int d^4x \sqrt{g} \left(-3\frac{k}{8\pi l^2}  
+ \frac{F^2}{16\pi} -{\cal V}(\Phi)  \right)
\end{equation}
which yields in, say, the $g=2$ case
\begin{equation}\label{prob}
P = \exp\left[ \left(2\pi\sigma r_s^2\sqrt{V(r_s)}\beta -\frac{q^2}{r_s r_+}(r_s-r_+)\beta
+ \frac{k(r_+^3-r_s^3)\beta}{\l^2}  - \frac{1}{8\pi\sigma^2}\right)
\right]
\end{equation}
for  the probability of pair creation of the black holes in the 
magnetic case; the electric case is similar but entails the incorporation 
of an additional surface term 
that vanishes in the magnetic case \cite{cosern}.  
Here $r_+$ is the location
of the outer horizon and $\beta$ the instanton period for the RNadS 
black hole. The above expression includes the case 
$k=1$ (RNdS); if the $\sigma$-dependent terms are omitted, and $r_s$ is taken to
be the location of the cosmological horizon, then the results of 
ref. \cite{cosern}
are recovered.  The probability (\ref{prob}) is relative to that for 
creation of a domain wall with no black holes or relativisitic 3-form.

To summarize, the C-metric has been shown to reduce to a set of metrics
whose Euclidean sections can be interpreted as instantons corresponding
to the pair creation of black holes with event horizons of arbitrary
topology.  The Lorentzian sections are spacetimes that are asymptotically
anti de Sitter and which can be interpreted as generalizations of
the $(2+1)$ dimensional black hole.  A more detailed study of these
black holes and their pair creation will appear in a forthcoming 
paper \cite{forth}.

\vspace{2cm}

\noindent{\bf Acknowledgements}

This work was supported by the Natural Sciences and Engineering
Research Council of Canada. I am grateful to J. Creighton for 
interesting discussions on this subject.


\begin{thebibliography}{99}
\bibitem{dgkt}H.F. Dowker, J.P. Gauntlett, D.A. Kastor and J. Traschen,
 Phys. Rev.  D {\bf 49}, 2909  (1994);
H.F. Dowker, J.P. Gauntlett, S.B. Giddings and G.T. Horowitz,
 Phys. Rev. D {\bf 50}, 2662 (1994);
S.W. Hawking, G.T. Horowitz and S.F. Ross, Phys. Rev. {\bf D51},
4302 (1995).
\bibitem{cosern}R.B. Mann and S.F. Ross, Phys. Rev. {\bf D52}, 2254
(1995).
\bibitem{strpair}S.W. Hawking and Simon F. Ross, Phys. Rev. 
Lett. {\bf 75} (1995) 3382;  R. Emparan Phys. Rev. Lett. {\bf 75} (1995) 3386 .
\bibitem{dompair}R.R. Caldwell, G.W. Gibbons, and A. Chamblin,
``Pair Creation of Black Holes by Domain Walls'', hep-th/9602216.
\bibitem{Cmetric}W. Kinnersley and M. Walker, Phys. Rev. D {\bf  2}, 
1359 (1970).
\bibitem{gen} J.F. Plebanski and M. Demianski, Ann. Phys. {\bf 98}, 98
(1976).
\bibitem{btz}M. Banados, C. Teitelboim and J. Zanelli,
Phys. Rev. Lett. {\bf 69}, (1992) 1849; M. Banados, M. Henneaux, C. Teitelboim and J. Zanelli,
 Phys. Rev. {\bf D} {\bf 48}, (1993) 1506.
\bibitem{amin}S. Aminneborg, I Bengtsson, S. Holst and P. Peldan,
``Making Anti de Sitter Black Holes'' gr-qc/9604022.
\bibitem{His}W.A. Hiscock  Phys. Rev. {\bf D35}, (1987) 1161.
\bibitem{Aur}A. Aurelia, R. Kissack, R.B. Mann and M. Spallucci,
  Phys. Rev. {\bf D35} (1987) 2961.
\bibitem{bjm}J.D. Brown and J.W. York, Phys. Rev. D {\bf 47}, 1420
(1993); J.D. Brown, J. Creighton and R.B. Mann, Phys. Rev. D {\bf 50}
6394 (1994). 
\bibitem{forth}R.B. Mann, in preparation.

\end{thebibliography}
\end{document}